\documentclass[preprint,12pt]{elsarticle}

\usepackage{graphics}

\usepackage{amssymb}

 \def\gsim{\mathrel{\rlap{\lower4pt\hbox{\hskip1pt$\sim$}}
 \raise1pt\hbox{$>$}}}

 \newcommand\la{\langle}
 \newcommand\ra{\rangle}
 \newcommand\beq{\begin{equation}}
 
 \newcommand\eeq{\end{equation}}
 \newcommand\beqn{\begin{eqnarray}}
 \newcommand\eeqn{\end{eqnarray}}
\def\mb{\,\mbox{mb}}
\def\fm{\,\mbox{fm}}

\def\GeV{\,\mbox{GeV}}
\def\TeV{\,\mbox{TeV}}
\def\lsim{\mathrel{\rlap{\lower4pt\hbox{\hskip1pt$\sim$}}
    \raise1pt\hbox{$<$}}}         
\def\gsim{\mathrel{\rlap{\lower4pt\hbox{\hskip1pt$\sim$}}
    \raise1pt\hbox{$>$}}}         
\def\J{J/\Psi}

\begin{document}

\begin{frontmatter}

\title{Nuclear suppression of $\J$: from RHIC to the LHC}

\author{B.\ Z.~Kopeliovich, I.\ K.~Potashnikova and Iv\'an Schmidt}

\address{Departamento de F\'{\i}sica
Universidad T\'ecnica Federico Santa Mar\'{\i}a; and
\\
Instituto de Estudios Avanzados en Ciencias e Ingenier\'{\i}a; and\\
Centro Cient\'ifico-Tecnol\'ogico de Valpara\'iso;\\
Casilla 110-V, Valpara\'iso, Chile}

\begin{abstract}
A parameter-free calculation for $\J$ suppression in $pA$ collisions, based on the dipole description, is confronted with the new data from the PHENIX experiment. Achieving good agreement, we employed this model predicting the contribution of initial state interactions (ISI) to $\J$ suppression in $AA$ collisions. Such a transition from $pA$ to $AA$ is not straightforward, since involves specific effects of double color filtering and boosting of the saturation scale.
Relying on this refined ISI contribution, we updated the previous analysis of RHIC data on $\J$ production in $Cu$-$Cu$ and  $Au$-$Au$ collisions at $\sqrt{s}=200\GeV$, and determined the transport coefficient of the created dense medium at $\hat q_0=0.6\GeV^2/\fm$. Nuclear effects for $\J$ production at the LHC are predicted using the transport coefficient $\hat q_0=0.8\GeV^2/\fm$, extracted from data on suppression of high-$p_T$ hadrons
in central lead-lead collisions at $\sqrt{s}=2.76\TeV$. Our analysis covers only direct $\J$ production, while data may also include the feed-down from decay of heavier states and $B$-mesons.

\end{abstract}


\end{frontmatter}


\section{Introduction}

Charmonium production in heavy ion collisions originates from two sources, which can be labeled as initial (ISI) and final (FSI) state interactions. The former are also called sometimes cold nuclear matter effects, and are related to the creation
of a $\bar cc$ pair and its subsequent propagation through the fast moving (in the collision c. m.) matter, consisting of nucleons and their remnants. The second stage, FSI, is related to the propagation and attenuation of the $\J$ through the (probably) dense matter created in the collision. This stage is usually the main point of interest, since it is considered as a probe for the properties of the created medium.

The two stages of collision are characterized by very different time scales, so no interference between them is possible. 
It has been always an important and challenging task to separate the effects originated from different stages of interaction. Essential information about the ISI stage can be withdrawn from data on $\J$ production in $pA$ collisions, where it cannot be mixed up with the effects of FSI. Although a model-independent extrapolation of ISI from $pA$ to $AA$ collisions is impossible \cite{nontrivial}, this is a good playground for the models.

In Section~\ref{pA} we perform parameter free calculations for $\J$ production in $pA$ collisions at $\sqrt{s}=200\GeV$ and different rapidities. The results agree well with the recently released data from the PHENIX experiment at RHIC. Predictions for the LHC energies are provided as well.

In Section~\ref{isi} we extend the model to heavy ion collisions, including the specific effects of double color filtering and boosting of the saturation scales. These effects do not exist in $pA$ collisions. The $p_T$-dependence of the nuclear ratio $R_{AA}$ is predicted at the energies of RHIC and LHC. This first stage of calculations cannot yet be compared to any data, since the effects of FSI are to be added.

in Section~\ref{fsi} we calculate the additional suppression factor related to attenuation of the produced $\J$ due to FSI with the created medium. The key observations here are the short formation time of the $\J$ wave function, and the relation between the break-up cross section of a $\bar cc$ dipole and the transport coefficient of the medium. If the latter is not known, the calculations cannot be done in a parameter-free way, but can be compared with data 
attempting to determine the transport coefficient. We found the parameter $\hat q_0=0.6\GeV^2/\fm$ for gold-gold collisions at $\sqrt{s}=200\GeV$. For the energies of LHC we rely on the analysis of the data on high-$p_T$ hadron suppression in lead-lead collisions at $\sqrt{s}=2.76\TeV$, which resulted in $\hat q_0=0.8\GeV^2/\fm$ and provided predictions for $\J$ suppression. 

Our results are displayed as function of $p_T$, as well as $p_T$-integrated.

\section{$\J$ production in $pA$ collisions}
\label{pA}

Any time scale is subject to Lorentz transformation and at high energies it may be long. In particular, the time scale for charm quark production, which is very short, $t^*_c\sim1/m_c$ in the proper reference frame, may become longer than the nuclear radius  in the nuclear rest frame,
\beq
t_c=\frac{2E_{\J}}{m_{\J}^2}\gg R_A.
\label{100}
\eeq
This condition is well satisfied at $\sqrt{s}=200\GeV$, where $l_c=12\fm$ at $y=0$, and rises as $e^y$ at forward rapidities. Therefore, a $\bar cc$ pair is not produced momentarily inside the nucleus, but it propagates through the whole nucleus. So, inclusion of the break up of the produced $\bar cc$ dipole, as is usually done, is not correct. Notice that this should not be a debatable issue, if one wants to comply with Lorentz invariance.

The long life time of a $\bar cc$ in the projectile hadron significantly enhances nuclear suppression. One can see that on the example of a very heavy nucleus. If $\J$ were suppressed only by inelastic processes, leading to break up of the $\bar cc$ dipole produced on mass shell, the $A$-dependence of the cross section would be $A^{2/3}$.
If however, the $\bar cc$ pair propagated through the whole nucleus, the $A$-dependence would be $A^{1/3}$.

The break up cross section for a $\bar cc$ dipole is usually assumed to be unknown and is  fitted to the same data which is to be explained. However, this cross section is well known from precise measurements at HERA. The proton structure function $F_2(x,Q^2)$, measured in a wide range of Bjorken $x$ and virtuality $Q^2$, provides a sensitive probe for the dipole (break up) cross section, dependent on the dipole size and energy.
Many parametrizations of the dipole break up cross section are currently available in the literature. One of the most popular results from a global analysis of HERA data gives a break up cross section parametrized in the saturated shape \cite{gbw}. In what follows we rely on this result.

In the regime of $t_c\gg R_A$, when the dipole size is "frozen" during propagation through the nucleus,
the nuclear suppression factor was calculated in \cite{nontrivial,kth-psi} as,
\beq 
R_{pA}={1\over A}\int d^2b\int\limits_{-\infty}^{\infty}dz\,
\left|S_{pA}(b,z)\right|^2,
\label{120}
\eeq
where
\beq
S_{pA}(b,z)=\int d^2r_T\,W_{\bar cc}(r_T)\,
\exp\left[-{1\over2}\sigma_{\bar ccg}(r_T)T_-(b,z)
-{1\over2}\sigma_{\bar cc}(r_T)T_+(b,z)\right].
\label{140}
\eeq
Here $W_{\bar cc}(r_T)\propto K_0(m_c r_T)\,r_T^2\,\Psi_{\J}(r_T)$ is the distribution function for the dipole size $r_T$;
$\sigma_{\bar ccg}(r_T)={9\over4}\sigma_{\bar cc}(r_T/2)-{1\over8}\sigma_{\bar cc}(r_T)$
is a three-body ($g\bar cc$) dipole cross section;
$T_-(b,z)=\int_{-\infty}^z dz'\rho_A(b,z')$;~ $T_+(b,z)=T_A(b)-T_-(b,z)$, and $T_A(b)=T_-(b,\infty)$. 

We employ the phenomenological dipole cross section $\sigma_{\bar cc}(r_T,x)$ fitted to HERA data in \cite{gbw}.  
Here 
\beq
x=\frac{\sqrt{\la M_{\bar cc}^2\ra+\la p_T^2\ra}}{\sqrt{s}}\,
e^{-y},
\label{160}
\eeq
where $y$ is the rapidity, and the mean invariant $\bar cc$ mass squared was fixed at the value corresponding to the color singlet model, $\la M_{\bar cc}^2\ra=2M_{\J}^2$.

\begin{figure}[htb]
\begin{center}
\includegraphics[width=8cm]{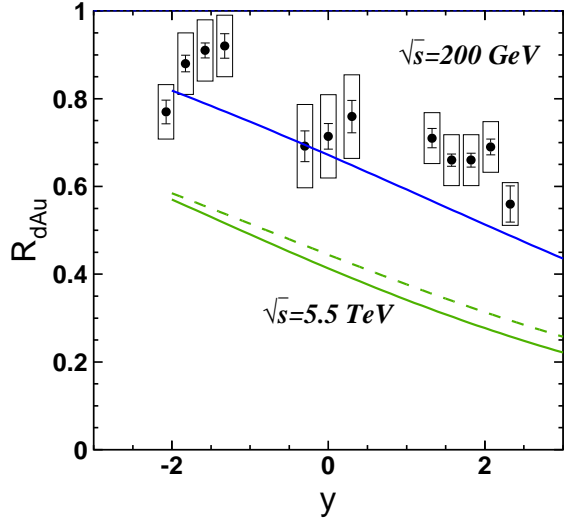}
\end{center}
\caption{\label{dAu} Data \cite{phenix-last} for the $p_T$-integrated nuclear suppression factor $R_{dAu}(y)$   for $\J$ produced in $d$-$Au$ collisions with rapidity $y$ at $\sqrt{s}=200\GeV$. The upper solid curve presents the result of Eqs.~(\ref{120})-(\ref{140}). It includes the small correction for gluon shadowing \cite{nontrivial}.
The lower solid (dashed) curve shows predictions for LHC at 
$\sqrt{s}=5.5\TeV$, including (excluding) gluon shadowing.}
 \end{figure}

Here we improve the accuracy of the calculations of \cite{nontrivial}, relying on a more realistic evaluation of the $\J$ mean radius, $\la r_T^2 \ra_{\J}={2\over3}\,\la r_{\J}^2\ra$, where $\sqrt{\la r_{\J}^2\ra}=0.42\fm$ \cite{buchmuller,hikt}. In Fig.~\ref{dAu} we compare the results with the new data \cite{phenix-last} for $\J$ suppression in deuteron-gold collisions at $\sqrt{s}=200\GeV$, and provide predictions for LHC.
 This calculation includes gluon shadowing, based on the global NLO analysis of DIS data \cite{ds} (see details in \cite{nontrivial}), although the correction is quite small.
The agreement with data is rather good,  taking into account that this is not a usual fit to the data to be explained, but is the first parameter free calculation.   Notice that the extension of our calculations down to $y<-1$ should be taken with precaution,  since the conditions for using the frozen dipole approximation, as well as dipole phenomenology itself, are not fulfilled.                         

We also calculated the nuclear modification of the $p_T$-dependence, known as Cronin effect. Following \cite{psi-AA,psi-bnl} we introduced into the integral over $\vec b$ in Eq.~(\ref{120}) a $p_T$-dependent factor, normalized to unity, 
\beq
R_{pA}(p_T,b)=\frac{\la p_T^2\ra\,R_{pA}}{\la p_T^2\ra+\Delta_{pA}(b)}
\left(1+\frac{p_T^2}{6\la p_T^2\ra}\right)^6
\left(1+\frac{p_T^2}{6[\la p_T^2\ra+\Delta_{pA}(b)]}\right)^{-6}.
\label{165}
\eeq
Here $\la p_T^2\ra$ is the mean transverse momentum squared for $\J$ produced in $pp$ collisions.
For the energy dependence we use the parametrization from \cite{psi-pp}, $\la p_T^2\ra=[-2.4+0.6\,\ln{s}]\GeV^2$.
$\Delta_{pA}(b)=\la p_T^2\ra_{pA}-\la p_T^2\ra_{pp}$ is nuclear broadening for a $\J$ produced at impact parameter $b$. It was calculated in \cite{broadening} in good agreement with the available data for $\J$ broadening,
\beq
\Delta_{pA}(b)={9\over4}\,C_q(E_{\J})\,T_A(b),
\label{170}
\eeq
where the factor $C_q(E_{\J})$ was calculated in \cite{broadening}.

\begin{figure}[htb]
\begin{center}
\includegraphics[width=8cm]{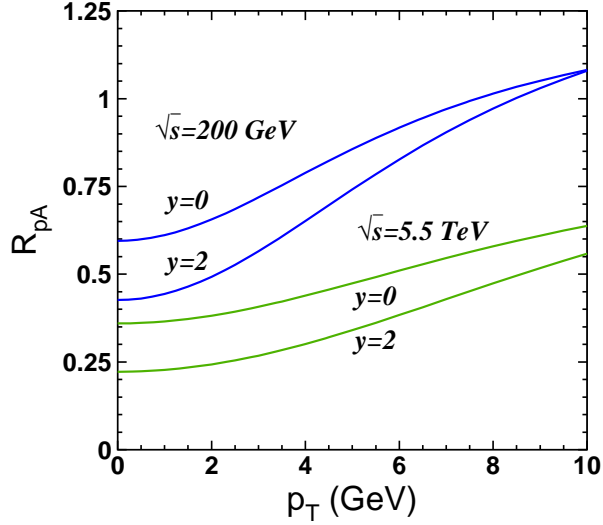}
\end{center}
\caption{\label{pA-pt} Prediction for the $p_T$-distributions of $\J$ produced with rapidities $y=0,2$ in $p$-$Au$ collisions at $\sqrt{s}=200\GeV$ and in $p$-$Pb$ collisions at $\sqrt{s}=5.5\TeV$.}
 \end{figure}

The results of the calculation of the nuclear ratio $R_{pA}(p_T)$ are depicted in Fig.~\ref{pA-pt} as function of $p_T$, at rapidities $y=0,2$ and at $\sqrt{s}=200\GeV$ and $5.5\TeV$.

Notice that these predictions are not reliable above $p_T\gsim5\GeV$, where data for $pp$ collisions are lacking, and the $p_T$ dependent factor Eq.~(\ref{165}) cannot be justified. An alternative shape of the $p_T$ dependence, suggested in \cite{psi-AA}, leads to similar results at $p_T\lsim 5GeV$, but a smaller Cronin effect at larger $p_T$.

\section{ISI effects in nuclear collisions}
\label{isi}

With this model for the nuclear effects related to ISI, tested in $pA$ collisions, one can move on to heavy ion collisions, and predict their ISI effects, which is desperately needed in order to single out from data the FSI suppression caused by the dense medium produced in the collision.

As was pointed out in \cite{nontrivial} the transition from $pA$ to $AA$ collisions is not straightforward and is model dependent. There are several effects specific for nucleus-nucleus collisions, and one of them is double color filtering.
Namely, color filtering of $\bar cc$ dipoles propagating through the first nucleus makes the mean dipole separation smaller, so the second nucleus becomes more transparent compared to attenuation of a $\bar cc$ dipole produced in $pA$ collisions. 
The ISI suppression factor in a collision of nuclei $A$ and $B$, including double color filtering was derived in \cite{nontrivial},
\beqn
R_{AB}(\vec b,\vec\tau)=
\frac{1}
{(\Lambda_A^+-\Lambda_A^-)(\Lambda_B^+-\Lambda_B^-)}\,
\ln\left[\frac{(1+\Lambda_A^-+\Lambda_B^+)(1+\Lambda_A^++\Lambda_B^-)}
{(1+\Lambda_A^++\Lambda_B^+)(1+\Lambda_A^-+\Lambda_B^-)}\right]
\label{180}
\eeqn
where $\vec b$ and $\vec\tau$ are the impact parameters of the collision and of the produced $\J$ respectively.
The other notations are,
\beqn
\Lambda_A^+&=&{\la r_T^2\ra\over2\,r_0^2(x_A)}\,\sigma_0T_A(\tau);
\label{200a}\\
\Lambda_A^-&=&{7\la r_T^2\ra\over32\,r_0^2(x_A)}\,\sigma_0T_A(\tau);
\label{200b}\\
\Lambda_B^+&=&{\la r_T^2\ra\over2\,r_0^2(x_B)}\,\sigma_0T_B(\vec b-\vec\tau);
\label{200c}\\
\Lambda_B^-&=&{7\la r_T^2\ra\over32\,r_0^2(x_B)}\,\sigma_0T_B(\vec b-\vec\tau).
\label{200d}
\eeqn
As was mentioned above, we rely on the parametrization of the dipole cross section \cite{gbw}, with parameters $\sigma_0=23.03\mb$ and $r_0(x)=0.4\times(x/x_0)^{0.144}\fm$, and $x_0=3.04\times10^{-4}$.
The value of $x_A$ is given by Eq.~(\ref{160}), and $x_B=x_A\,e^{2y}$. The mean separation squared $\la r_T^2\ra$ of the produced $\bar cc$ dipole was evaluated in \cite{nontrivial} at $\la r_T^2\ra=0.045\fm^2$.

\begin{figure}[htb]
\begin{center}
\includegraphics[width=8cm]{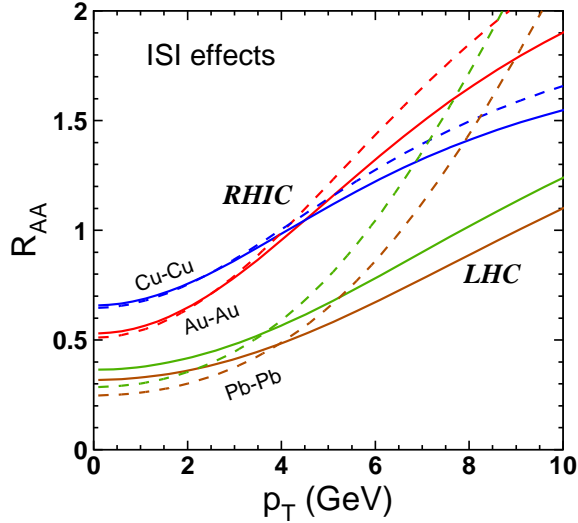}
\end{center}
\caption{\label{isi-pt} Nuclear ratio for $\J$ produced in central ($b=0$) $AA$ collisions including only the ISI and excluding FSI effects. Solid lines correspond to (from the top to bottom) $Cu$-$Cu$ and $Au$-$Au$ at $\sqrt{s}=200\GeV$, $Pb$-$Pb$ at $\sqrt{s}=2.76$ and $5.5\TeV$. The dashed curves are different from the solid ones by inclusion of the effect of boosted saturation scale (see text for explanations). }
 \end{figure}

The ISI stage also modifies the $p_T$ distribution of produced $\J$s, as was discussed in the previous section.
The full effect of ISI, including attenuation and the modified $p_T$ dependence, has the form 
of the ratio Eq.~(\ref{180}) multiplied by the normalized factor,  Eq.~(\ref{165}), with the replacement $\Delta_{pA}(b)\Rightarrow \Delta_{AB}(\vec b,\vec\tau)=\Delta_{pA}(\vec\tau)+\Delta_{pB}(\vec b-\vec\tau)$.
The $p_T$-dependent nuclear ratio for $\J$ produced with $y=0$ in central $Cu$-$Cu$ and $Au$-$Au$ collisions at $\sqrt{s}=200\GeV$; and in $Pb$-$Pb$ at $\sqrt{s}=2.76$ and $5.5\TeV$, is depicted in Fig.~\ref{isi-pt}. These calculations include all the ISI effects, but exclude FSI.

We also calculated the effects of boosting the saturation scale in nuclear collisions \cite{boosting}, which leads to an increase of broadening $\Delta_{AA}(\vec b,\vec\tau)$. The enhancement factor $K_{AA}(\vec b,\vec\tau)$ was calculated for $\J$ production in \cite{nontrivial}.
We plotted the results by dashed curves, since this effect is still under debate and has not been confirmed by data so far.

\section{FSI suppression and the transport coefficient}
\label{fsi}

\begin{figure}[htb]
\begin{center}
\includegraphics[width=8cm]{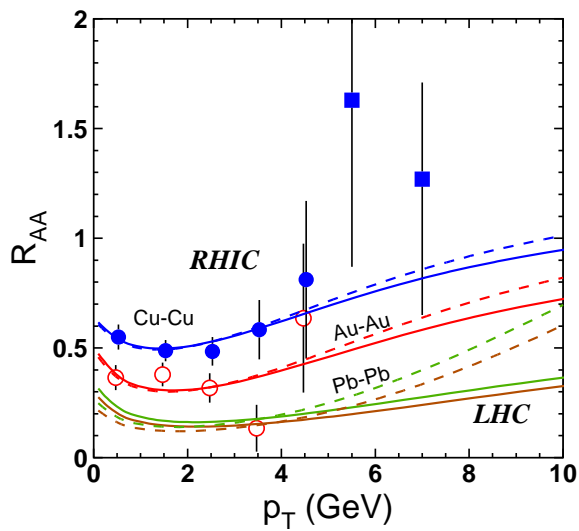}
\end{center}
\caption{\label{AA-pt} RHIC data \cite{phenix1,phenix2,star} for the nuclear modification factor $R_{AA}$ for $\J$ production in central collisions of $Cu$-$Cu$ (closed circles and squares) and $Au$-$Au$ (open circles) at $y=0$ and $\sqrt{s}=200\GeV$.
The curves are the same as in Fig.~\ref{isi-pt}, but corrected for the FSI effects, which are calculated with the transport coefficient for gold-gold $\hat q_0=0.6\GeV^2/\fm$ adjusted to the data. $R_{AA}$ at the LHC energies is predicted with $\hat q_0=0.8\GeV^2/\fm$, extracted in \cite{high-pt} from data for nuclear quenching of high-$p_T$ hadrons \cite{alice}. }
 \end{figure}

The nuclear effects related to ISI shown in Fig.~\ref{AA-pt} present the results of parameter-free calculations, which are, however, model dependent. Although the model was tested in Section~\ref{pA} comparing with $d$-$A$ data, the transition from $pA$ to $AA$ collisions is very nontrivial \cite{nontrivial} and involves more modeling, as was discussed in the previous section. 
Nevertheless, relying on these results for the ISI contribution, one can try to pin down the effects of FSI, which probe the properties of the medium created in nuclear collisions. Such an analysis of RHIC data for $\J$ production in heavy ion collisions at RHIC was performed in \cite{psi-AA,psi-bnl}. Here we update the analysis relying on a more refined calculation of the ISI effects, and also make predictions for the LHC.

The key point of the analysis \cite{psi-AA} for the FSI stage is the relation between the dipole cross section and broadening \cite{dhk,jkt}, which allows to express $\J$ attenuation in terms of the transport coefficient of the medium,
\beq 
S(L)=\exp\left[-{1\over3}\,\la r_{\J}^2\ra
\int\limits_0^L dl\, \hat q(l)\right]. 
\label{240} 
\eeq
An important observation also employed here, is the shortness of the formation time of the $\J$ wave function. This is why the transverse $\bar cc$ dipole separation squared is fixed at ${2\over3}\la r_{\J}^2\ra$.
 
The transport coefficient $\hat q$ is defined as the magnitude of broadening experienced by a quark propagating through
a path length $1\fm$ in the medium \cite{bdmps}. With the usual assumptions that the initial medium density at $t=t_0$ is proportional to the number  of participants $n_{part}$, and the density is diluting with time as $1/t$,
we arrive at
\beq 
\hat q(t,\vec b,\vec\tau)=\frac{\hat q_0\,t_0}{t}\, 
\frac{n_{part}(\vec b,\vec\tau)}{n_{part}(0,0)},
\label{260} 
\eeq
where the parameter $\hat q_0$ is the maximal value of $\hat q$, for the medium produced at $t=t_0$ in central collision at $b=\tau=0$. As previously, $\vec b$ and $\vec\tau$ are the impact parameters of collision and position of the quark, which is propagating through the medium in the transverse direction (in the collision c.m.).

Combining  the $\J$ attenuation factor caused by FSI, Eq.~(\ref{240}), with the ISI effects depicted in Fig.~\ref{isi-pt},
as is described in \cite{psi-AA,psi-bnl}, we arrive at the final nuclear modification of the $\J$ production rate, which can be now compared with data, as is shown in Fig.~\ref{AA-pt}.
The parameter $\hat q_0=0.6\GeV^2/\fm$ for gold-cold collisions was adjusted to RHIC data \cite{phenix1,phenix2,star}, shown by solid and open circles for $Cu$-$Cu$ and  $Au$-$Au$ respectively. This parameter was rescaled for copper-copper according to
Eq.~(\ref{260}) as $(\hat q_0)_{Cu}=[n_{part}^{Cu}(0,0)/n_{part}^{Au}(0,0)](\hat q_0)_{Au}=
0.38\GeV^2/\fm$.

In order to predict $R_{AA}$ at the energies of LHC, one needs to know the corresponding values of $\hat q_0$.
This parameter was found to be $\hat q_0=0.8\GeV^2/\fm$ in the analysis \cite{high-pt} of the recently published data \cite{alice} on suppression of high-$p_T$ hadrons 
in central lead-lead collisions at $y=0$ and $\sqrt{s}=2.76\TeV$. We use this value of $\hat q_0$ at both energies 
$\sqrt{s}=2.76$ and $5.5\TeV$, since do not expect a significant variation of $\hat q_0$ in this energy range.
The results for the LHC are depicted in Fig.~\ref{AA-pt} by the bottom pair of solid curves.

Dotted curves show the effect of mutual boosting of the saturated scales in colliding nuclei. We see that at $\sqrt{s}=200\GeV$ this correction is quite small and is visible only at $p_T>5\GeV$, where calculations, as was mentioned above, are not reliable anyway. At the energies of the LHC the correction is larger and may be detected if data were sufficiently precise.

We also calculated the nuclear modification factor for the $\J$ production cross section integrated over $p_T$ at $y=0$ and different energies.
The results are presented in Table~\ref{table1}.
\begin{table}[hb]
\caption{The $p_T$-integrated nuclear ratio for $\J$ production at $y=0$ in central collisions
of different nuclei at the energies of RHIC and LHC.}
\begin{center}
\begin{tabular}{|c|c|c|}
\hline
$Cu$-$Cu$& $\sqrt{s}=200\GeV$ & $R_{AA}=0.54$\\[0mm]
$Au$-$Au$&$\sqrt{s}=200\GeV$ & $R_{AA}=0.34$\\[0mm]
$Pb$-$Pb$&$\sqrt{s}=2.76\TeV$ & $R_{AA}=0.18$\\[0mm]
$Pb$-$Pb$&$\sqrt{s}=5.50\TeV$ & $R_{AA}=0.16$\\[0mm]
\hline
\end{tabular}
\end{center}
\label{table1}
\end{table}

\section{Summary}

Summarizing, we tested parameter-free calculations for $\J$ suppression in $pA$ collisions, based on the dipole description \cite{kth-psi,nontrivial}, with the new data from the PHENIX experiment \cite{phenix-last}. Encouraged by the good agreement, we employed this model predicting the ISI contribution to $\J$ production in $AA$ collisions. Such a transition from $pA$ to $AA$ involves novel effects of double color filtering and boosting the saturation scale \cite{nontrivial}, which were included.
The previous analysis \cite{psi-AA} of RHIC data on $\J$ production in $Cu$-$Cu$ and  $Au$-$Au$ collisions at $\sqrt{s}=200\GeV$ is updated, and  the transport coefficient of the created dense medium is found to be $\hat q_0=0.6\GeV^2/\fm$. The estimated accuracy of this analysis is about $30\%$.
To predict nuclear effects for $\J$ production in $Pb$-$Pb$ at the LHC we rely on the value of the transport coefficient $\hat q_0=0.8\GeV^2/\fm$, determined from the analysis of data on suppression of high-$p_T$ hadron production
in central lead-lead collisions at $\sqrt{s}=2.76\TeV$ \cite{alice}. 

Finally, a word of caution is in order: the calculations are done for direct production of $\J$, while currently available data may be significantly contaminated with a feed-down from decays of $\Psi^\prime$, $\chi$, and $B$-mesons. 
This may significantly disturb the magnitude of the observed nuclear effects, because heavy charmonia apparently are more suppressed by FSI than $\J$, while $B$-mesons are expected to be much less affected by the ISI and FSI effects \cite{kt-heavy-q}.
\bigskip

{\bf Acknowledgments: }
We are thankful to Will Brooks for encouraging us to perform these calculations, 
and to Jamie Nagle for providing data tables for nuclear effects in $d$-$Au$ collisions.
This work was supported in part by Fondecyt (Chile) grants 1090236,
1090291 and 1100287,  by DFG (Germany) grant PI182/3-1, and by
Conicyt-DFG grant No. 084-2009.

\bibliographystyle{model1-num-names}



\end{document}